# Who Wrote this?
# How Smart Replies Impact Language and Agency in the Workplace

Manuscript

Author:              Kilian Wenker
                     kilian.wenker@fau.de

Version date:        2023-02-17

Revised copy (several minor corrections, revised subsections 2.2-2.4)

## Highlights

- The loss of agency theory states that the use of AI leads to a loss of human agency
- Smart replies, a text-based form of artificial intelligence, have users respond to stimuli
- Those responses substantiate a transfer between human and machine agency through priming and time pressure
- Smart replies influence the content we author and the behavior we generate
- Human agency transfers out faster than previously thought, but can be enhanced at the same time

*Article*

# Who Wrote this? How Smart Replies Impact Language and Agency in the Workplace

Kilian Wenker

Faculty of Business, Economics, and Law, Friedrich-Alexander-Universität Erlangen-Nürnberg; 91054 Erlangen, Germany; kilian.wenker@fau.de

**Abstract:** AI-mediated communication is designed to help us do our work more quickly and efficiently. But does it come at a cost? This study uses smart replies to show how AI influences humans without any intent on the part of the developer—the very use of AI is sufficient. I propose a loss of agency theory as a viable approach for studying the impact of AI on human agency. This theory focusses on the transfer of agency that is forced by circumstances (such as time pressure), human weaknesses (such as complacency), and conceptual priming. Mixed methods involving a crowdsourced experiment test that theory. The quantitative results reveal that machine agency affects the content we author and the behavior we generate. But it is a non-zero-sum game. The transfers between human and machine agency are fluid; they complement, replace, and reinforce each other at the same time.

**Keywords:** Human Agency, Machine Agency, Human-AI Interaction, Priming, Time Pressure, Computer-Mediated Communication (CMC), Social Heuristics, Smart Replies, Linguistic Alignment, Sentiment

## 1. Introduction

Smart replies (SRs) are short responses suggested by artificial intelligence (AI) software that allow you to quickly reply to incoming messages. They afford mainly saving time and minimizing interruptions, which makes them particularly valuable in business contexts. SRs are designed to help us do our work more quickly and efficiently. But do they come at a cost?

Human agency in this article is a choice and decision-making capacity, including the corresponding ability to act, i.e., both power of decision and power of action (there is a lack of consensus on the definition of the term, see [1]). A recent synthesis of key research themes and trends in AI-mediated communication (AI-MC) states that users retain "substantial agency: they choose which suggested message to use or to ignore, and may also modify the message" [2] (p. 92). However, suggested texts can affect us in various ways without us being aware of it. For instance, lack of time in day-to-day business, a tendency to rely too heavily on automation, or complacency can guide us. When such circumstances prevent the user from thinking at length about the best possible response, they de facto limit the user's ability to retain agency. Since this transfer of agency is forced by circumstances, loss of agency would be a more appropriate term.

Loss of agency, even if compelled by circumstances, can still be deliberate. But there may also be subliminal stimuli that blur this boundary of deliberate choice. Priming occurs when previous SR suggestions determine responses because the SR user would have responded differently if they had not received those suggestions in the first place [3–5].

In other words, that allegedly substantial degree of agency may be much smaller than we presume. This has huge implications. AI assistance could be manipulative without any intention behind it. As written communication in the form of brief texts has become a ubiquitous means of communication, this argument becomes even more compelling. AI might not only be disruptive as a technology, but also for the meta discourse on communication theories.

Although researchers have drawn attention to SRs (e.g., [2, 5–10]), empirical studies are scarce and limited in scope. Moreover, they do not directly investigate manipulative effects through priming or time constraints.

Because AI is not a single, discernable thing, a brief review and precise definition of SRs is warranted. This leads to research question *RQ1: What defines SRs?* Next, I seek to explore the framework for biases in SRs, and ask *RQ2: How do SRs impact language?* Last, I focus on the construct of agency and raise *RQ3: To what extent do SR users transfer agency?* Table 1 outlines the research questions and the contributions of this article.

**Table 1.** Research questions and contributions

| Research Question | Contribution |
|---|---|
| RQ1: What defines SRs? | I analyze the literature to date and derive a genus-differentia definition of SRs. |
| RQ2: How do SRs impact language? | I give an account of the linguistic constraints imposed by allowlists, blocklists, filters, canned responses, and various biases in the training data as well as in response generation (e.g., shorter, simpler wording or the avoidance of emotion). Those agency constraints at the linguistic level form the basis for action stimuli. |
| RQ3: To what extent do SR users transfer agency? | I propose a theory of loss of agency inspired by the relevant literature on computer-mediated communication (CMC), AI-MC, and SRs. Toward creating an empirical foundation for the theory, I combine a crowdsourced experiment and interviews. Evidence suggests that the use of SRs can lead to a loss of agency in the digital workplace (DWP) through response priming and time pressure. |

Section 2 introduces related work, proposes a theory on loss of agency and forms corresponding hypotheses. Section 3 presents the experiment design, its operationalization, and its sample. Section 4 addresses the empirical findings. Section 5 concludes the work.

**2. Background**

*2.1 Smart replies and the digital workplace*

CMC refers to interpersonal communication transmitted over digital devices. When CMC becomes agentic through AI, e.g., through the feature of generating content, as is the case with SRs, CMC becomes AI-MC. Figure 1 depicts a characterization of SRs within CMC. The rounded boxes within the SRs frame indicate the NLP methods that SRs use.

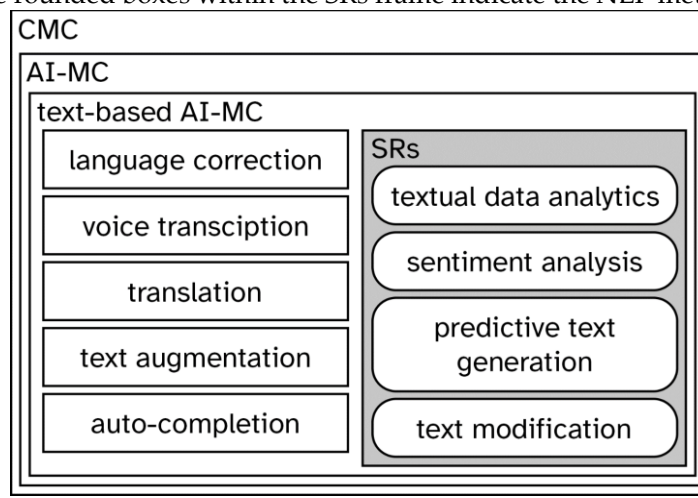

**Figure 1** Smart replies within CMC

The term smart reply is ambiguous: On the one hand, there is Google's "Smart Reply" functionality, a widely discussed implementation of SRs. On the other hand, SRs are an academic term describing a type of AI-mediated communication (see the use of the term e.g., in [2, 6, 11–14]). In this article, Google's Smart Reply is capitalized to distinguish it from the general term.

Table 2 provides a summary of nine sources with descriptions of the various aspects of SRs.

**Table 2.** Features, outcomes, and context for SRs

| Aspect of SRs | Sources | Comment |
|---|---|---|
| SRs are an example of AI-MC | [2, 5, 9, 15, 16] | Genus proximum et differentia specifica is a recognized definitional approach. However, differentia specifica needs to be elaborated. |
| Human-human text conversation with AI support | [2, 5, 17]; [6] mention this in their introduction; [15] characterize this as "positioned somewhere between human-human and human-smart agent communication" (p. 16) | While this is true, it is redundant if one assumes that AI-MC is by definition interpersonal communication. Typically, AI-MC is implicitly defined as interpersonal communication by defining it as a subset of CMC. |
| Text-based | [5, 6] | Unclear terms like various messaging apps [15], examples like chat, text, and email [9] or Gmail, Outlook, Skype, Facebook Messenger, Microsoft Teams, and Uber [10] do not explicitly indicate "text-based" media |
| Suggestions | [5–7, 9, 15, 17] | Most authors agree that SRs are presented to the user in the form of suggestions. Ying et al. [10] resort to the more general term "assist", which could also encompass other AI-MC mechanisms. |
| Reply suggestions | [6, 7, 9, 10, 15, 17] | SRs are always phrased in reaction to an incoming message. This is exclusively about pre-written answer texts, not autocorrection or formulation assistance. |
| Short suggestions | [7, 9, 10, 15] | SRs do not come in the form of long texts. They usually range from a single word to short sentences. |
| Complete and plausible suggestions | [5, 7, 10]; [7] stress and explain "plausible" | These are obvious but crucial features of SRs. "Plausible" is the only attribute that refers to the content of SRs. |
| Save time | quick [6, 10, 15] one tap [7, 9] reduce keystrokes [9] | Most authors explicitly state that SRs save users' time. There is a risk of conflation between purpose and effect. Regardless, this is a key component of the concept and therefore needs to be included in the definition. |
| Underlying technology | AI according to [2, 5–7, 9, 11–13] | The authors agree that AI is the underlying technology. Since AI is included in the definition of AI-MC, I will not explicitly mention the underlying technology to avoid redundancy. |
| Mobile device(s) | Mobile (phone) according to [7]; [9] mention the role of the interface (e.g., mobile, desktop, tablet); [6] list desktops, tablets, and mobile devices | Restricting SRs to mobile devices does not make sense, because the functionality and use do not depend on the device, albeit SRs are mostly seen on mobile devices. If SRs are used on a desktop, they are still SRs. Following the principle of parsimony, I will not include this item in the definition. |

Based on these extant definitions and descriptions, I propose a genus-differentia definition of SRs as *a form of AI-MC that offers short, complete and plausible reply suggestions in a text-based communication, allowing the user to save time*. This definition emphasizes several points. First, SRs are not an algorithm or an underlying technology, but a concept. Second, SRs are an offer and its users must react to AI stimuli. SRs initiate human action. That makes them popular in burgeoning AI-MC research. Third, SRs predict plausible responses, requiring the system to have some kind of interpretive ability and foresight. Later on, we will see that SRs are a prime example of machine agency and that the latter can enhance human agency.

SRs are a general-purpose tool for a wide range of use cases. A notable exception is the use of SRs in One-Click-Chat, the in-app chat of the mobility service provider Uber. Here, the operational context is specific and the list of possible SRs is domain specific. A typical situation during pickup would be a driver receiving an incoming rider message asking: "Where are you right now?" [14] (p. 2597).

**The digital workplace**

While the workplace has traditionally been understood as the physical place where work was performed, a distinctive feature of the DWP is precisely that it is possible to work in a wide variety of locations at any time, as digitization enables mobile workplaces. Throughout this article, the term DWP will refer to an integrated technology platform that provides all the tools and services that enable employees to do their jobs effectively, both alone and with others, regardless of location [18]. Thus, smartphones can be part of the workplace or even represent the entire DWP.

Note that this definition is truncated. The original definition includes that the DWP "is strategically coordinated and managed through DWP designs that are agile and capable of meeting future organizational needs and technologies" [18] (p. 480). I suppress this part because it is partly circular (DWP is something managed through DWP designs) and partly not conclusively ascertainable (future-proof might or might not exclude certain DWP technologies because they will one day be superseded by a new technology).

The branched arrow in Figure 2 containing two converging paths shows how the user deploys the AI-MC to send a SR to a receiver. The lower path shows how the user does not use the SR suggestions and sends a message directly, without using SR technology. The upper path shows the route the message takes when a SR suggestion is tapped (or clicked) and the wording is generated by the AI. Although this decision is assumed to be deliberate, controlled, and signed off, in this case the recipient is effectively talking to an AI impersonating the user.

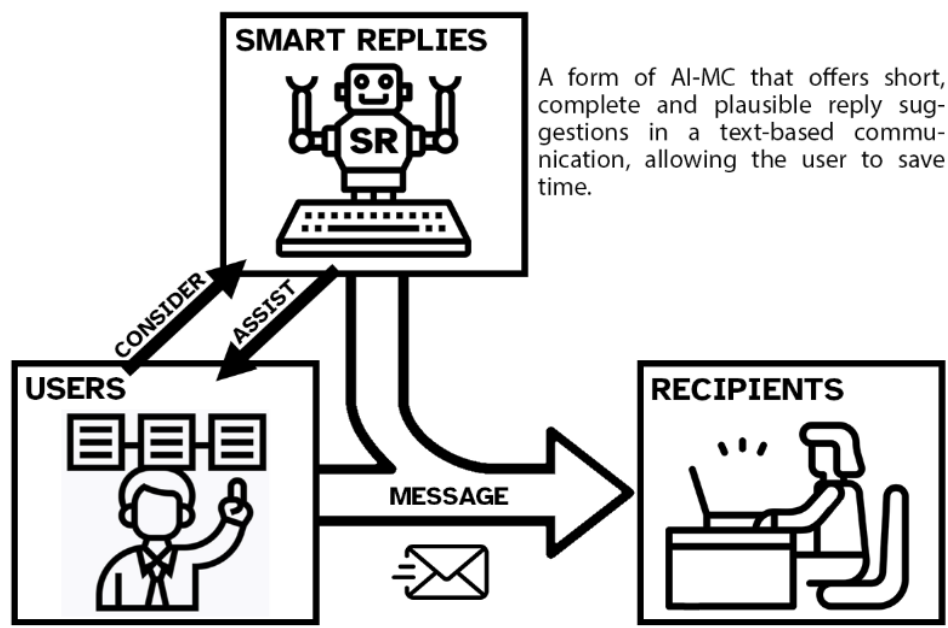

**Figure 2** Smart replies in the digital workplace

*2.2 Impact on language*

We live in a biased world. "Applying machine learning to ordinary human language results in human-like semantic biases" [19] (p. 183). Yet AI-MC may be purposely coded to encourage users to modify communications that bear the mark of prejudice. SRs may avoid some biases, e.g., gender bias by blocklisting gender-specific pronouns. But new biases may be created as well, e.g., shorter and simpler sentences. For instance, Google researchers found that common responses with high prior likelihood appeared more often and adapted the score system [12]. Table 3 shows that difference using sample emails.

**Table 1** Examples of SR suggestions with and without response bias

| Message: Did you print the document? | |
|---|---|
| **With response bias** | **Without response bias** |
| – Yes, I did. | – It's printed. |
| – Yes, it's done | – I have printed it. |
| – No, I didn't | – Yes, all done. |

Source: [12] (p. 8)

There are at least two biases in the left column. The wording is more generic, consistent with recent findings that people tend to choose shorter, simpler formulations when writing with suggestions [12, 20]. And one negative answer is now offered, see e.g., the rules "no more than two suggestions should be from the same semantic group" [13] (para. 9), "greater semantic variability and intrinsic diversity" [11] (p. 1), or "semantic intent clustering" [7] (p. 959).

The longer the reply is, the more important it is to the user that the reply is phrased in their own tone (Gawley as cited in [21]). At six words it is still acceptable to sound generic, but by ten words it becomes awkward [21]. Regardless of the word count, users are reluctant to let AI talk about emotional or otherwise intimate matters (Gawley as cited in [21]). Therefore, Smart Reply has filters on emotional statements, because "the early iterations of Smart Reply were overly affectionate. 'I love you' was the machine's most common suggested response. This was a touch awkward: because the model has no knowledge of the relationship between an email's sender and its receiver, it provides the same suggested responses whether you are corresponding with your boss or a long-lost sibling" [21] (para. 6). Evidence suggests that AI-generated responses change the expression of emotion in human conversations, raising concerns that we may lose our personal communication style as language becomes more homogeneous over time [6].

Prior research implies positivity biases in SRs [5, 17]. However, experimental control over which suggestions were presented does not necessarily prove bias. It is also conceivable that after training the models, users' preferences for short, positive, and polite responses are accurately represented and reflect implicit values. "Languages, English included, appear to be inherently positive" [9] (p. 3). The team that developed the Smart Reply functionality argues that a positive attitude reflects the style of email conversations: Positive replies may be more common, and when negative responses are warranted, users may prefer less direct wording [7]. Their model forces negative suggestions when at least one positive suggestion is included and none of the top three replies are negative [7]. The issue is that we do not (yet) know where to set the typical level of positive language in a given situation.

SRs could lead to an ultimately limited vocabulary. For example, Google's Smart Reply response options are limited to "a relatively small human-curated whitelist with limited context-dependency" [22] (p. 2288). Personalization of SRs could help remedy this in the future. However, this will be a long journey. For instance, when users of a predictive writing app emailed that they would like to meet with an investor the following week, the experimental models suggested meeting with him, but when the user entered the same wording about a nurse, the experimental models suggested meeting with her [22]. To limit the impact, developers removed all suggestions with a gender pronoun [22], which is—while great against gender bias—an obvious vocabulary restriction. Another example is LinkedIn, an employment-oriented social network, adding controls to ensure that profanity is not suggested in their SRs for member messages [13].

Overall, language is modified by algorithms that pay attention to word count (of reply suggestions), politeness (e.g., filters against profanity), gender sensitivity (e.g., no gender pronouns), response selection (e.g., forcing a variety of semantic clusters; tending to suppress emotionality), simplicity (more generic wordings; canned responses), probably positivity, and, most importantly, biased training data and often predefined response sets.

The modified language of SRs has implications for human language production. We are thinking in words. The interactive alignment model [23] argues that interlocutors become more linguistically aligned over the course of a conversation. Some researchers already pondered the limiting effects on creativity in writing [6, 20, 24]. This reasoning entails that by using SRs, our human-produced language would gradually adapt to the linguistic features of SRs, i.e., we would unconsciously adjust our terminology, maybe formulate shorter sentences, perhaps lose certain linguistic nuances, etc.

In addition to the general impact on language, could there be impacts on individual choices? A study addressing biases in predictive typing on mobile devices debated whether system biases might influence what people create and argues that the sentiment could be affected because "it is easy to enter via accepting the recommendation verbatim" [3] (p. 34). It found evidence that people were primed by the system's recommendations [3]. A laboratory experiment concluded that compound human-AI messages contained "marginally more positive emotion words" (p. 7) compared to the control condition without SRs [5]. However, writers often compensate for overly positive responses by reducing positive sentiment [9].

In sum, prior work shows that biases in training data lead to SR biases, but it is not yet clear if biased SRs in turn lead to biased human behavior.

*2.3 CMC theories and machine agency*

Early CMC theories pounced on the missing paralinguistic cues of CMC (e.g., media richness theory, social presence theory). In sum, those theories argue that CMC is inferior to face-to-face communication due to the loss of nonverbal cues. This applies to SRs: text-based communication inherently lacks paralinguistic cues.

Anthropomorphic theories such as the CASA paradigm (computers are social actors) describe the tendency of people to subconsciously attribute human characteristics, feelings, or intentions to non-human entities [25–28]. While early CMC theories focus on the limitations of CMC, anthropomorphic theories expose the shortcomings of the human mind. Especially interactions involving verbal prompts automatically elicit social responses and, in sum, "whenever there is discourse, people assume there is an underlying subject who speaks" [4] (p. 27). Anthropomorphism may occur, e.g., when a SR user talks to their smartphone and berates it for not making enough suggestions.

Later theories acknowledge the merits and affordances of CMC and stress that CMC constraints do not necessarily impair communication. Media synchronicity theory states that what matters is not the richness of a medium but its synchronicity, i.e., the extent to which people work on the same task simultaneously [29], implying that the benefit of SRs might be to keep the dialogue going with short response times (immediacy of feedback).

Compensatory adaptation theory (CAT) takes this a step further, stating that interlocutors can compensate or even overcompensate for the obstacles posed by CMC through procedural structuring [30]. For example, interlocutors could compensate for the text-based character of the communication by putting more cognitive effort into writing, or by being clearer and more concise. While the SR user has a limited choice of response options and therefore cannot perform procedural structuring, their AI could be coded to do so.

There is a nascent fourth wave of CMC theories as AI increasingly removes agency from human users. This school of thought will form AI-MC theories and examine the advent of machine agency (sometimes designated as AI agency, digital agency, or agentic artifact).

The machine agency paradigm expands the concept of agency to automation, assuming that AI decides and acts in a self-determined manner. Thus, agency is no longer exclusively a human capacity. Machine agency asserts goal-directed behavior and actions not directly controlled by a human [2, 15, 16, 19, 28, 31, 32]. Some denote this as a capability to act autonomously [31, 33]. That self-determined character may be challenged by the fact that decisions are made by an AI simulation rather than a conscious mind: algorithms determine machine agency, so that the latter is more of a technically pre-programmed ability to decide and act. Although agency is sometimes understood as the ability to think or reason, and the sense of agency sometimes refers to the sense of being in charge (i.e., that our own actions do not simply happen to us), these definitional approaches are not the focus of this study. The role of technology has progressed from functional tools to agentic co-players, and this study focusses on how this progress affects human agency. We do not debate whether a non-conscious entity can have agency at all, but rather examine whether humans confer agency on an agentic artifact in an uncontrolled way.

Relying on a tool does not necessarily mean that you surrender your agency to the tool. For example, relying on a simple tool like an alarm clock does not mean that I surrender my agency to the alarm clock. But if I use a sophisticated alarm clock that selects the wake-up time according to a complicated algorithm to achieve optimal sleep, then I transfer the power of decision (and thereby agency) to the alarm clock in a controlled way. In contrast to the normal alarm clock scenario, I do not know the exact outcome here, yet I trust that the machine agent will make a good enough choice. However, we are interested in whether and how relying on a tool can lead to an uncontrolled transfer of human agency.

*2.4 Loss of agency theory and research hypotheses*

I propose a theory on the loss of human agency caused by the use of SRs or, in a generic way, by the use of AI. This theory focusses on the transfer of agency that is forced by circumstances (such as time pressure), human weaknesses (such as complacency), and psychological priming. The theory illustrates that the use of SRs cannot be separated from the contextual contingencies in which SRs are used. The term "loss" is intended to reflect that the transfer of agency is not controlled by SR users.

Machine agency is inherent in SRs because the algorithm makes suggestions on its own. That does not yet mean that agency is transferred from the human user to the machine agent. Alas, the interactions between human agency, machine agency, and other constructs are too complex, too convoluted, too unexplored to be fully addressed in this study. The basic assumption here is that human agency and machine agency are not exogenous variables imposed on the model, but are interrelated in ways that augment and replace each other. Agentic artifacts shape their users.

One of those other constructs mentioned above is *priming*, which draws on a stimulus-response scheme. In conceptual priming, the meaning of the input stimulus, i.e., the meaning of the prime (e.g., a word, a sentence, a SR) impacts the target (e.g., the response message). For example, if an employee is asked whether they would attend a meeting and they receive overly positive SR suggestions, they may subconsciously consider that to be the most appropriate response. When constructs such as priming limit an individual's ability to take decisions based on careful consideration, they result in a partial, involuntary give-off of decision-making power, in our case to a machine agency. Such a loss of power of decision can hardly be controlled by the user of SRs, since priming is a nonconscious process. Recall that power of decision is per definition one of the two essential dimensions of agency.

Further elements are not necessarily unconscious factors, but often factors of which we are not actively aware at the moment. Overtrust in AI (e.g., in the form of complacency) or situational constraints (e.g., time pressure) may prevent the SR user from thinking at length about the best possible action and thus result in another loss of agency.

*Complacency* refers to insufficient vigilance or vetting of AI. Complacency denotes that a SR user may not care about inaccurate answers suggested to them by SR technology because they are focusing their attention on other issues in a multitask environment. *Automation bias* (AB) refers to the use of decision support systems as a shortcut tool, often in conjunction with a favorable attitude toward automated decision-making, ignoring conflicting information. Both terms coin the tendency to shed responsibility for communication, either as an attention allocation strategy or as a tendency to over-rely on automation (see [34] for a review). A laboratory experiment [35] found that AB "is the result of using automation as a heuristic replacement for more vigilant and complete information search and processing" (p.713), i.e., the issue might be less a fundamental belief in the relative authority of automated aids, but rather that "automation is used as a decision-making short cut that prematurely shuts down situation assessment" (p. 714).

*Situational constraints* such as lack of time in day-to-day business, a high mental workload, uncertainty etc. often force shortcuts [36–38]. When time is perceived to be short, cognitive processing either speeds up [39] or becomes less likely and individuals tend to rely on heuristics [36], owing to an inherent contradiction between response speed and decision accuracy [37].

Taken together, I hypothesize:

- **H1:** The SR options will include more positive sentiment than negative or neutral sentiment.
- **H2:** The use of SRs will lead to a loss of human agency through priming.
- **H3:** When SRs are used, situational constraints such as lack of time will lead to a loss of agency.

Note that H1 is hypothesis H1a from [5].

In broader terms, the very use of AI leads to a loss of agency. The model in Figure 3 shows the relationships between the various constructs on a theoretical plane that led to hypotheses H2 and H3 at the empirical level of research.

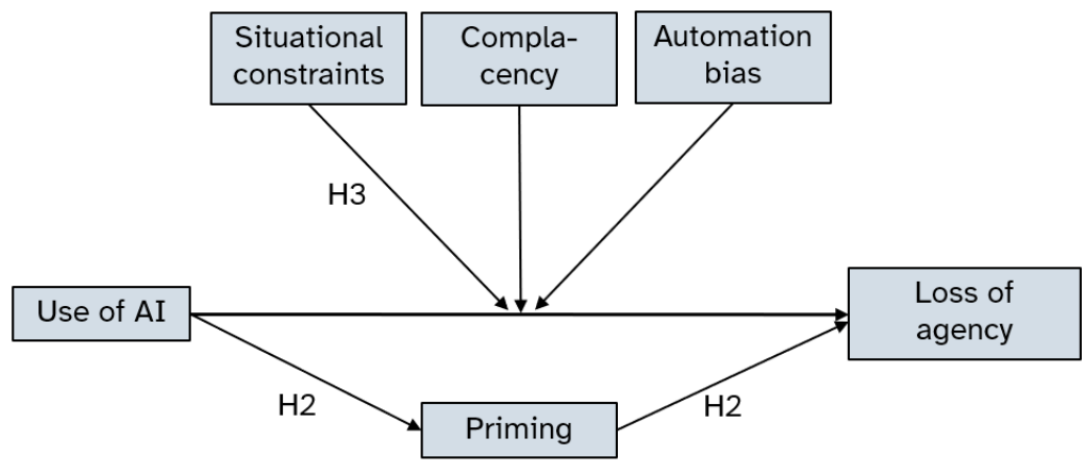

**Figure 3** A loss of agency model

Arrows show the influence of variables on each other. The point in the middle of Figure 3 where three arrows converge does not represent a local point or moment, but simply shows the effect relationships of moderating variables on the dependent variable. Situational constraints, complacency, and automation bias are moderating variables. Priming is a mediating variable.

I do not form hypotheses regarding complacency and AB because I could not test them in the experiment (Sections 3 and 4). I was unable to do this because the works councils of the two companies involved did not agree to questions that might relate to personal work performance.

## 3. Experiment

*3.1 Experiment design*

Employees of two large companies (manufacturing and software) in Germany were asked online to respond to twelve emails, using SRs, or alternatively to formulate their own answers. Survey data in the companies had to be collected via internal company platforms due to internal data protection guidelines. The time period of the study was July 2022.

The head of sales of the manufacturing company sent me the content of his last 20 emails, completely anonymized. I shortened these emails so that survey respondents would not abandon the survey due to time constraints. Then I sent these emails to a newly created email account and noted the SR suggestions: 14 emails were offered SRs (see e.g., [7] for an explanation of filtering out messages that are not eligible for SRs). Twelve of these e-mails were not duplicates in terms of content. The survey containing these twelve emails and their matching SRs is hereafter referred to as AI-SRs.

For this study, two groups of SR suggestions are needed to compare their effects. In a second step, I asked five employees of the software company to formulate three short, typical responses to the twelve email texts from their perspective. I then grouped the responses semantically and selected the three responses with the most occurrences to obtain three human-written SRs per email, henceforth referred to as H-SRs. They served as a control group, although two suggested responses in email 5 written by humans were identical to those generated by the AI.

Another design would have been to pre-sort the AI-generated SRs according to sentiment and then split them into two groups. However, AI suggestions were mostly positive, ranging from 0.5 to 0.9. A third option would have been to make no suggestions at all to the control group (see [5]). However, this would have left no possibility to measure the sentiment of the SR suggestions in the control group (no suggestions means no measure of the offered sentiments).

In sum, participants were randomly offered one of three surveys:
- **H-SRs** Human-written SR suggestions for the twelve emails.
- **AI-SRs** AI-generated SR suggestions for the twelve emails.
- **AI-SRs+TC** AI-SRs, but with a time constraint of a few seconds to respond to each of the twelve emails.

The AI-SRs+TC setting was probably closest to real life in the workplace and was used to simulate the effects of the independent variable of situational constraints.

Response options were always reshuffled to avoid bias due to the order of the proposed SRs.

Participants were instructed to read the emails carefully and respond according to their own assessment, in case of doubt or uncertainty with answers that should be neutral to moderately agreeing. All three groups had the option to compose their own response text for each email. An open-ended question was subsequently used to inquire about the participants' experiences and assessment regarding SRs. The survey concluded with a brief demographic section (gender and age group).

*3.2 Measures*

The construct of agency has to be measured indirectly, since we cannot look into each other's minds. What we can measure is a skewing of decisions, which implies an impact on the SR author's agency (as power of decision). More specifically, I drew on the sentiment of written responses, whether entered through SR selection or manual typing, for two reasons. First, meaning is at the core of conceptual priming, and sentiment is an important part of the intention or meaning of a business message or SR. Second, sentiment may be a better measure of agency than a subjective "sense of agency" in the context of priming. The sense of human agency may be very high when human agency is actually not high—recall that priming is a nonconscious process. I used the analysis feature of Google Cloud NLP API which is available for German. This use of Google's tool to analyze sentiment in order to gauge priming is identical to the approach used by [3] for the same aim.

The entire response, i.e., mostly a short sentence, was used to evaluate sentiment. Individual words are less appropriate here [3]. Other SR researchers used Mechanical Turk workers [15, 17], two research assistants and the dictionary-based classifications of VADER and LIWC [5], or VADER combined with word shifts [9].

The independent variable of situational constraints was simulated by tight time constraints, i.e., a few seconds of process time. Its effects can be measured by the gap in sentiment of given replies between the groups AI-SRs+TC and AI-SRs.

*3.3 Sample*

Participants were recruited via emails sent by the head of human resources at a software company and the sales director at a manufacturing company. Participants received no compensation. The sample consisted of 346 participants. All participants wrote German. Of the participants, 57.2% (n=198) identified themselves as women and 42.8% (n=148) identified themselves as men, all preferred to disclose their gender. 54.6% of participants worked in the manufacturing company, 45.4% in the service company. One participant was excluded from analysis due to technical problems.

The age distribution was collected in cohorts 18-24, 25-34, 35-44, 45-54, 55-64, and 65 years and older. Each cohort had a 20-24% share, except for the 55-64 cohort, which had an 11% share; no participants were in the 65 and older cohort.

To determine whether nonresponse bias was present, I compared early responders with late responders. No significant differences were found. However, one respondent phrased their own response for each email and indicated in the open-ended section of the questionnaire that they considered SRs fundamentally inappropriate in the corporate context.

All participants were debriefed about the purpose of the experiment after completion.

**4. Results and discussion**

*4.1. Results*

4.1.1. Quantitative results

Consistent with prior work [5, 17] and with H1, sentiment analysis found 86% of AI-SR options included positive sentiment (and 14% negative sentiment). There was a positivity bias in H-SRs as well, but to a lesser extent (78% included positive sentiment, 5% neutral sentiment, and 17% negative sentiment).

Figure 4 shows the average sentiment scores for the surveys H-SRs, AI-SRs and AI-SRs+TC. Google Cloud NLP calculates sentiment analysis values in a range from -1.0 for maximally negative to +1.0 for maximally positive. The initial values of the arrows indicate the average sentiment scores of the suggested SRs. The end of the arrows indicates the average sentiment scores of the given answers.

Without priming by the SR suggestions, the values on the right side of Figure 4 should be at the same level. Instead, the more positive SRs suggestions in AI-SRs (score 0.55) with otherwise unchanged conditions compared to the suggested H-SRs (score 0.36), lead to higher sentiment scores of the given responses (0.58 AI-SRs and 0.64 in AI-SRs+TC compared to 0.47 in H-SRs).

Figure 4 shows that the offered sentiment successfully created a difference in sentiment valence between the responses given by the groups H-SRs and AI-SRs. Writers given more positive recommendations accepted or included more positive content.

Tukey's all-pairs HSD test confirmed the difference in sentiment means between all three groups was significant at the p=0.01 level (3 groups; df = 342; critical level 4.12): $q_{tukey}$ = 18.89 for H-SRs and AI-SRs+TC, $q_{tukey}$ = 13.05 for H-SRs and AI-SRs, and finally the lowest but still significant, $q_{tukey}$ = 5.84 for AI-SRs and AI-SRs+TC.

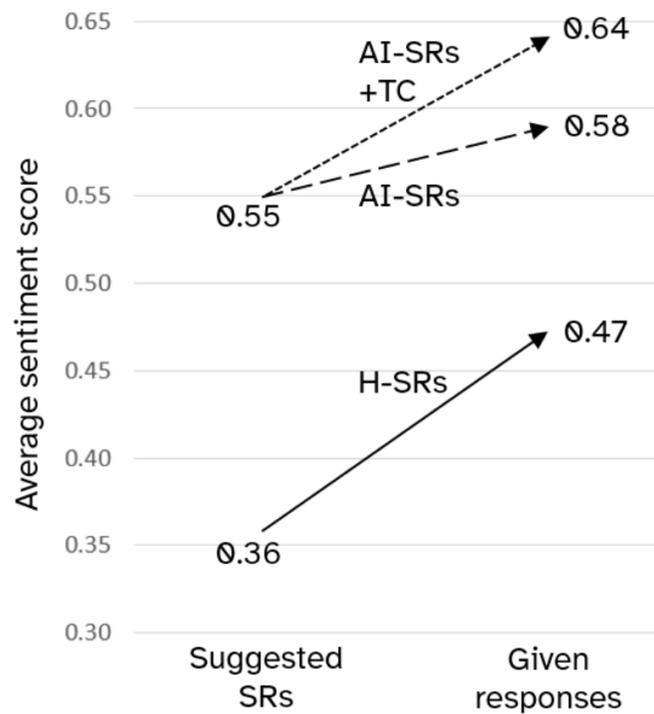

**Figure 4** Sentiments scores of suggested SRs and all given responses

Next, I zoomed in on the acceptance rates for each email (Table 4 and Figure 5).

**Table 4** SR acceptance rates

|  | H-SRs | AI-SRs | AI-SRs+TC |
|---|---|---|---|
| All emails | 92.1% | 87.4% | 83.0% |
| Email 6 | 88.9% | 66.6% | 59.4% |
| Email 7 | 91.6% | 90.9% | 72.9% |
| Email 8 | 83.3% | 69.6% | 54.0% |

The fact that H-SRs have the highest acceptance rate on average can be explained by the better quality of human-written SRs. However, this does not explain why the acceptance rate does not increase under time pressure, but decreases (see first row).

A detailed breakdown of acceptance rates, i.e., e-mail by e-mail, revealed that emails 7 and especially 6 and 8 sorted out the wheat from the chaff. Email 6 asked for confirmation of arrival including time, and here participants were less likely to accept the generally worded SRs suggestions and preferred responding with a specific time. Email 7 is a more complex email that indirectly asks a second question. Email 8 directly asks two questions in one body of text. It is noteworthy that participants in the AI-SRs+TC group skipped the suggested SRs 46.0% of the time with email 8.

Figure 5 depicts the acceptance rates of the two groups AI-SRs and AI-SRs+TC. These two groups are best suited to gauge the effect of time pressure: priming by SR suggestions should not differ here because the stimuli are exactly the same, and the only difference is no time pressure versus time pressure.

The vertical axis indicates the average acceptance rate for the SRs offered. The horizontal axis is an ordinal scale of the twelve emails, sorted by average rate of the two groups in declining order.

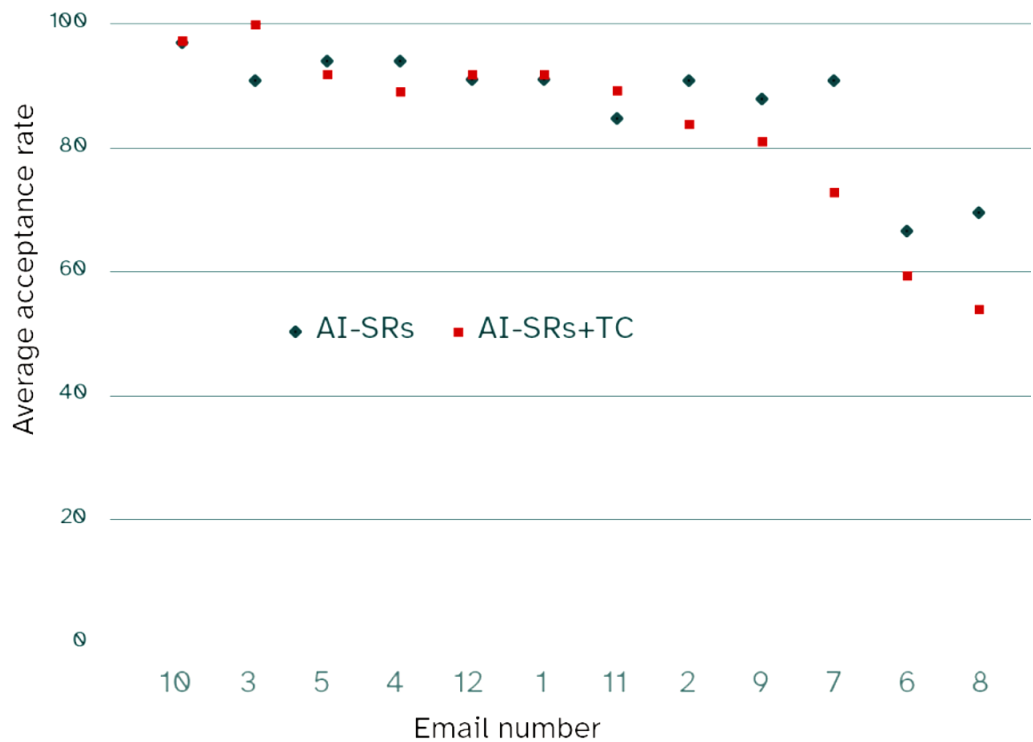

**Figure 5** Breakdown of acceptance rates by email

When the quality of the suggested SR was of satisfactory grade, the SR suggestions were accepted under time pressure at roughly similar rates compared to the group without time pressure (AI-SRs). In one case, the AI-SRs+TC group has a 100% acceptance rate: email 3, a note of thanks for documents sent, which was most often returned with "You're welcome! "

However, Figure 5 visualizes that the acceptance rate has a threshold above 80% that, when crossed downward, causes the acceptance rate in the AI-SRs+TC group to drop faster. If we assume that the acceptance rate is a quality parameter, then—below a certain threshold—the behavior changes under time pressure. SR users in a time crunch who find that the SR suggestions may not really fit their intended message do not spend time thinking about the SR suggestions, but quickly write the answer themselves. This is especially true for faster typists (see [20]).

This is noteworthy because it indicates that some participants resisted time pressure and loss of agency. The AI-SRs+TC group manifested two effects. First, there was response priming compared to the AI-SRs group (although less significant than in the pair H-SRs versus AI-SRs). And second, the tolerance for semi-good responses decreased.

The acceptance rate for H-SRs (not shown in Figure 5) is generally higher and falls off more slowly than for AI-SRs.

4.1.2 Participant experience

Most participants pointed to the time savings, describing their impression as a "quick response on the smartphone where you don't want to type for a long time" [P9]. Shorter response times were also frequently mentioned: "You're more likely to give immediate feedback because you have to spend a lot less time" [P328].

Participants saw potential usefulness such as "assistance in wording responses" [P172], "no spelling errors in short answers" [P302], "positive impact on the customer (a simple "thank you" may be enough to increase sympathy)" [P65], "facilitation for non-native speakers" [P25], "faster and friendlier relationship management" [P124]. One participant appreciated that there were "no meaningless phrases" [P15], maybe due to the rule of diverse semantic clusters.

While some participants noted that the SR suggestions tended to be generic or "standardized" [P335] or only useful for "trivial answers" [P60], others echoed the statements of CAT: "clearer and more targeted answers" [P204], "short and concrete information" [P15], or "faster answers, more politeness, fewer misunderstandings" [P162]. Some participants highlighted how SRs enhance human agency, e.g., text comprehension of the incoming email: "you know immediately what the other person means" [P279], or text composition of the outgoing email: "to process a mail immediately without having to switch to 'write mode' " [P324].

Participants also indicated why they did not use the suggested SRs: "more work, so no benefits. More like saving time typing" [P96], "quick answer, although the accuracy of the answer will probably be poor in many cases" [P301], to outright rejection: "time saving, you don't have to think about it, for lazy people. [...] I would nevertheless like to make it clear that I do not support this form of answering at all, it is completely impersonal, superficial and we are degenerating more and more into robots. At least a short, friendly individual reply will probably still bring everyone together …" [P35].

Participants also confirmed complacency and AB in their own behavior. Comments such as "you don't have to read the emails quite so carefully" [P267] illustrate complacency, "you don't have to think about answers yourself and quicker" [P92] illustrate AB as heuristic shortcut, "less thinking about the right answer because one of the given ones seems appropriate in terms of content and language" [P5] illustrate AB as confidence in AI, or "you don't have to think about the wording of the answer (whether it is too polite, too direct, etc.)" [P276] may illustrate both constructs. One comment summarized: "quicker response because you don't have to think about how to answer first" [P89].

Some participants deliberately counteracted the loss of agency by writing the wording of the suggested SR verbatim in the text box for self-formulated responses—sometimes by copying two SR proposals into one. This behavior was also observed in two other studies, see [P10] in [9], designated as "desire for greater control" (p. 12f.), and [24], designated as "chaining multiple suggestions" (p. 9).

Despite all linguistic filters, one participant limited the range of application in the DWP context: "among colleagues certainly a good idea for quick coordination. Towards customers and other 'unknowns' I don't find it appropriate" [P144]. The qualitative interviews explicate on this aspect.

4.1.3 Qualitative interviews

This study consisted of two parts, the crowdsourced experiment and the qualitative interviews, with only selected participants (N=8) participating in the latter. The qualitative interview results broadly mirror the quantitative results. There was a perceived trade-off between speed of response and accuracy of response (see [37] for the speed-accuracy tradeoff). Most refused trade-offs between speed of response and correctness of response (i.e., accuracy is so low that the response becomes incorrect) in a DWP context, e.g., "there is no compromise to be made on the correctness of the reply. If a SR does not reflect the correct content, I simply don't use the SR and write a manual reply" [I6], or "I refuse to send half-correct answers, no matter how big my time crunch is. In business, you break more than it's worth" [I3]. The heuristic that intervenes here is obviously that time pressure in business should not lead to hasty, wrong decisions. Experience, expertise, and situational awareness seem to be an excellent recipe against wrong decisions under time pressure.

One reason for the basic rejection of certain SRs in the workplace seemed to be the lack of formality in the wording of the proposed SRs—they seemed too informal.

*4.2. Discussion*

The experiment extends prior research on SRs and employs a novel set of human-written SRs as a control group because we do not know where to set the typical level of positivity (here in a German-language DWP context).

**Hypothesis H1** that SR suggestions would include more positive sentiment than negative or neutral sentiment is supported.

Evidence supports **hypothesis H2** as well. Email writers who received more positive SR suggestions wrote more positive content. Since sentiment is being used as a proxy measure for agency, evidence amounts to a loss of agency and that AI technology is impacting human responses already.

The results support **hypothesis H3**: Under time pressure, email writers increased their average positive sentiment compared to the group with the same SR suggestions but without time constraints. However, a threshold for the acceptance of SR suggestions in everyday business is observable (see Figure 5) and plausible (see participant experience): under time pressure, a well-fitting offer is accepted more quickly, but a poor SR suggestion is rejected more quickly and entails immediate typing. When weighing reaction speed against decision accuracy, participants basically opted for the latter.

Table 5 summarizes the findings.

**Table 5** Research questions and results

| Research question | Result |
|---|---|
| RQ1: What defines SRs? | SRs are a form of AI-MC that offers short, complete and plausible reply suggestions in a text-based communication, allowing the user to save time. |
| RQ2: How do SRs impact language in the workplace? | Linguistic constraints imposed by allowlists, blocklists, filters, canned responses, and various biases in the training data as well as in response generation, are inbuilt into action stimuli and can result in the language becoming more homogeneous over time. |
| RQ3: To what extent do users transfer agency? | Evidence suggests that the use of SRs lead to a loss of agency in the DWP through response priming, and under time pressure. |

The questionnaire allowed for termination at any time, but forced responses to the emails, either by clicking on the suggested response templates or by typing own responses. However, in some rare cases, participants created their own additional categories. One participant noted in the blank response box of email 11 that they would not write a response at all, but would call the customer. Another participant reported that they would not answer to email 3. One participant commented that there was no need to respond to email 1 because the customer would call anyway.

Note that we did not hit upon a simple give-off toward machine agency in the both constraining and affording relationship between human and machine agency. The trade-off is a non-zero-sum game because it has cooperative elements. If that trade-off were a completely competitive zero-sum game, any loss of human agency would instantly increase machine agency. But there are gains despite the loss. The use of AI can enhance human agency, e.g., through procedural structuring according to CAT, or immediacy of feedback according to media synchronicity theory, both confirmed by participant experience in subsection 4.1. Another example, albeit purely conceptual, would be to include location information in One-Click-Chat so that the driver can easily provide an estimate of arrival time and accurate distance information, precomposed by SR technology. In short, the trade-off is not strictly human versus machine agency, there is also human-AI teaming. We may need to understand the interaction as working together with machine agency in a team, i.e., as a synergy of collective agency, not as a sum—with all the pros and cons of working with good and bad team members.

Because this experiment required participants to write under artificial conditions, the biggest threat to the external validity of the results is that the situation described in the emails was not detailed enough or participants did not take the experiment seriously enough. On the other hand, the situations were taken from real life and are very common; many will receive such emails every day. Participants were asked by their supervisors to answer the emails as realistically as possible. The verbal comments of the participants and their actual behavior, such as high rejection rates for certain SRs, give good reason to assume that the participants' responses do not deviate significantly from the natural situation. My sample consisted of employees in Germany. The results might not generalize to other populations, i.e., external validity cannot be tested.

## 5. Conclusion

SRs constrain and deform language with algorithms that pay attention to word count, simplicity, politeness, gender, response selection, probably positivity, and, most importantly, biased training data and predefined response sets. In sum, those skews lead to SR biases (like positivity), and in a second step, this study investigated if those biased SRs in turn lead to biased human behavior.

Using agency as theoretical lens, we hit upon human behavior that was biased through response priming and time pressure. Time pressure had an additional effect: participants became pickier when SR quality fell below a threshold.

Technical advances such as, most recently, Chat-GPT, a language model optimized for dialogue that delivers high-quality responses, herald a more intense and ubiquitous adaptation of predictive writing tools that will, in turn, have a stronger impact on users. This is a clear indication that the larger conversation about machine agency is gaining momentum and will continue to do so in the coming years.

The loss of agency model and empirical evidence suggest that machine agency affects the content we author and thereby the behavior we generate. In coarse terms, this study investigates the trade-off in human "versus" machine agency. However, those transfers between human and machine agency are fluid; they complement, replace, and reinforce each other at the same time.

**Funding:** This research received no external funding.

**Conflicts of Interest:** The author declares no conflict of interest.

**Abbreviations**

The following abbreviations are used in this article:

| | |
|---|---|
| AI | artificial intelligence |
| AI-MC | AI-mediated communication |
| API | application programming interface |
| App | mobile application (software) |
| CAT | compensatory adaption theory |
| CMC | computer-mediated communication |
| CASA | computers are social actors |
| df | degree of freedom of the denominator |
| DWP | digital workplace |
| HCI | human-computer interaction |
| HSD | (Tukey's) honest significant difference |
| I | interviewee (in the participant interviews) |
| IS | information systems |
| ML | machine learning |
| N | total number of participants |
| n | sample size of a particular group |
| NLP | natural language processing |
| P | participant (of the experiment) |
| p | p-value |
| SR | smart reply |
| SRs | smart replies |

**References**

1. Engen, V., Pickering, J.B., Walland, P.: Machine Agency in Human-Machine Networks; Impacts and Trust Implications. In: Kurosu, M. (ed.) *Human-Computer Interaction. Novel User Experiences*. pp. 96–106. Springer International Publishing, Cham (2016). https://doi.org/10.1007/978-3-319-39513-5_9.
2. Hancock, J.T., Naaman, M., Levy, K.: AI-mediated communication: Definition, research agenda, and ethical considerations. *Journal of Computer-Mediated Communication*. 25, 89–100 (2020). https://doi.org/10/gj4mj4.
3. Arnold, K.C., Chauncey, K., Gajos, K.Z.: Sentiment bias in predictive text recommendations results in biased writing. In: *Proceedings of Graphics Interface*. pp. 33–40 (2018). https://doi.org/10.20380/GI2018.07.
4. Brahnam, S.: Building character for artificial conversational agents: Ethos, ethics, believability, and credibility. *PsychNology Journal*. 7, (2009).
5. Mieczkowski, H., Hancock, J.T., Naaman, M., Jung, M., Hohenstein, J.: AI-mediated communication: Language use and interpersonal effects in a referential communication task. Proc. ACM Hum.-Comput. Interact. 5, 1–14 (2021). https://doi.org/10/gp9p99.
6. Hohenstein, J., DiFranzo, D., Kizilcec, R.F., Aghajari, Z., Mieczkowski, H., Levy, K., Naaman, M., Hancock, J., Jung, M.: Artificial intelligence in communication impacts language and social relationships. arXiv:2102.05756 [cs]. (2021).
7. Kannan, A., Kurach, K., Ravi, S., Kaufmann, T., Tomkins, A., Miklos, B., Corrado, G., Lukacs, L., Ganea, M., Young, P., Ramavajjala, V.: Smart reply: Automated response suggestion for email. In: Proceedings of the 22nd ACM SIGKDD International Conference on Knowledge Discovery and Data Mining. pp. 955–964. ACM, San Francisco California USA (2016). https://doi.org/10/gp9p95.
8. Naeem, M.A., Linggawa, I.W.S., Mughal, A.A., Lutteroth, C., Weber, G.: A smart email client prototype for effective reuse of past replies. IEEE Access. 6, 69453–69471 (2018). https://doi.org/10/gp9p98.
9. Robertson, R.E., Olteanu, A., Diaz, F., Shokouhi, M., Bailey, P.: "I can't reply with that": Characterizing problematic email reply suggestions. In: Proceedings of the 2021 CHI Conference on Human Factors in Computing Systems. pp. 1–18. ACM, Yokohama Japan (2021). https://doi.org/10/gksmfp.
10. Ying, Q., Bajaj, P., Deb, B., Yang, Y., Wang, W., Lin, B., Shokouhi, M., Song, X., Yang, Y., Jiang, D.: Language scaling for universal suggested replies model. arXiv preprint arXiv:2106.02232. (2021).
11. Deb, B., Bailey, P., Shokouhi, M.: Diversifying reply suggestions using a matching-conditional variational autoencoder. arXiv preprint arXiv:1903.10630. (2019).
12. Henderson, M., Al-Rfou, R., Strope, B., Sung, Y.-H., Lukács, L., Guo, R., Kumar, S., Miklos, B., Kurzweil, R.: Efficient natural language response suggestion for smart reply. arXiv preprint arXiv:1705.00652. (2017).
13. Pasternack, J., Chakravarthi, N.: Building smart replies for member messages. Available online: https://engineering.linkedin.com/blog/2017/10/building-smart-replies-for-member-messages, accessed on 15/02/2023.
14. Weng, Y., Zheng, H., Bell, F., Tur, G.: OCC: A smart reply system for efficient in-app communications. In: Proceedings of the 25th ACM SIGKDD International Conference on Knowledge Discovery & Data Mining. pp. 2596–2603. ACM, Anchorage AK USA (2019). https://doi.org/10/gf7ndd.
15. Hohenstein, J., Jung, M.: AI as a moral crumple zone: The effects of AI-mediated communication on attribution and trust. *Computers in Human Behavior*. 106, 1–30 (2020). https://doi.org/10/gk49jg.
16. Jakesch, M., French, M., Ma, X., Hancock, J.T., Naaman, M.: AI-mediated communication: How the perception that profile text was written by AI affects trustworthiness. In: Proceedings of the 2019 CHI Conference on Human Factors in Computing Systems. pp. 1–13. ACM, Glasgow Scotland Uk (2019). https://doi.org/10/gjvmw4.
17. Hohenstein, J., Jung, M.: AI-supported messaging: An investigation of human-human text conversation with AI support. In: Extended Abstracts of the 2018 CHI Conference on Human Factors in Computing Systems. pp. 1–6. ACM, Montreal QC Canada (2018). https://doi.org/10/gp9p9v.
18. Williams, S.P., Schubert, P.: Designs for the digital workplace. *Procedia Computer Science*. 138, 478–485 (2018). https://doi.org/10/ghcpst.
19. Caliskan, A., Bryson, J.J., Narayanan, A.: Semantics derived automatically from language corpora contain human-like biases. Science. 356, 183–186 (2017). https://doi.org/10/f93cpf.
20. Arnold, K.C., Chauncey, K., Gajos, K.Z.: Predictive text encourages predictable writing. In: Proceedings of the 25th International Conference on Intelligent User Interfaces. pp. 128–138. ACM, Cagliari Italy (2020). https://doi.org/10/gp9p9w.
21. Twilley, N.: Google's new autoreply sounds great!!!! Available online: https://www.newyorker.com/tech/annals-of-technology/google-new-smart-reply-sounds-great, (2015) accessed on 15/02/2023.
22. Chen, M.X., Lee, B.N., Bansal, G., Cao, Y., Zhang, S., Lu, J., Tsay, J., Wang, Y., Dai, A.M., Chen, Z., Sohn, T., Wu, Y.: Gmail smart compose: Real-time assisted writing. In: Proceedings of the 25th ACM SIGKDD International Conference on Knowledge Discovery & Data Mining. pp. 2287–2295. ACM, Anchorage AK USA (2019). https://doi.org/10/gf7m2t.
23. Pickering, M.J., Garrod, S.: An integrated theory of language production and comprehension. *Behav Brain Sci*. 36, 329–347 (2013). https://doi.org/10/f45frn.
24. Buschek, D., Zürn, M., Eiband, M.: The impact of multiple parallel phrase suggestions on email input and composition behaviour of native and non-native English writers. In: Proceedings of the 2021 CHI Conference on Human Factors in Computing Systems. pp. 1–13. ACM, Yokohama Japan (2021). https://doi.org/10/gksk4z.

25. Kim, J., Merrill, K., Xu, K., Sellnow, D.D.: I like my relational machine teacher: An AI instructor's communication styles and social presence in online education. *International Journal of Human–Computer Interaction*. 37, 1760–1770 (2021). https://doi.org/10.1080/10447318.2021.1908671.
26. Nass, C., Steuer, J., Tauber, E.R.: Computers are social actors. In: Proceedings of the SIGCHI conference on Human factors in computing systems. pp. 72–78 (1994).
27. Schuetzler, R.M., Grimes, G.M., Scott Giboney, J.: The impact of chatbot conversational skill on engagement and perceived humanness. *Journal of Management Information Systems*. 37, 875–900 (2020). https://doi.org/10/ghrmzz.
28. Westerman, D., Edwards, A.P., Edwards, C., Luo, Z., Spence, P.R.: I-It, I-Thou, I-Robot: The perceived humanness of AI in human-machine communication. *Communication Studies*. 71, 393–408 (2020). https://doi.org/10/gjvf3f.
29. Dennis, A.R., Valacich, J.S.: Rethinking media richness: Towards a theory of media synchronicity. In: Proceedings of the 32nd Annual Hawaii International Conference on Systems Sciences. 1999. HICSS-32. Abstracts and CD-ROM of Full Papers. pp. 10-pp. IEEE (1999). https://doi.org/10/bj2bqh.
30. Kock, N., Lynn, G.S., Dow, K.E., Akgün, A.E.: Team adaptation to electronic communication media: Evidence of compensatory adaptation in new product development teams. *European Journal of Information Systems*. 15, 331–341 (2006). https://doi.org/10/df7838.
31. Baird, A., Maruping, L.M.: The next generation of research on IS use: A theoretical framework of delegation to and from agentic IS artifacts. MISQ. 45, 315–341 (2021). https://doi.org/10/gmpdk3.
32. Sundar, S.S.: Rise of machine agency: A framework for studying the psychology of human–AI interaction (HAII). *Journal of Computer-Mediated Communication*. 25, 74–88 (2020). https://doi.org/10/ggjvvq.
33. Ågerfalk, P.J.: Artificial intelligence as digital agency. *European Journal of Information Systems*. 29, 1–8 (2020). https://doi.org/10/gjfm6g.
34. Goddard, K., Roudsari, A., Wyatt, J.C.: Automation bias: a systematic review of frequency, effect mediators, and mitigators. *J Am Med Inform Assoc*. 19, 121–127 (2012). https://doi.org/10/fndc86.
35. Skitka, L.J., Mosier, K., Burdick, M.D.: Accountability and automation bias. *International Journal of Human-Computer Studies*. 52, 701–717 (2000). https://doi.org/10.1006/ijhc.1999.0349.
36. Burmeister, C.P., Moskaliuk, J., Cress, U.: Ubiquitous working: Do work versus non-work environments affect decision-making and concentration? *Frontiers in Psychology*. 9, 1–11 (2018). https://doi.org/10/gqptt3.
37. Heitz, R.P.: The speed-accuracy tradeoff: history, physiology, methodology, and behavior. *Frontiers in Neuroscience*. 8, (2014). https://doi.org/10/gfw8p2.
38. Peters, L.H., O'Connor, E.J.: Situational constraints and work outcomes: The influences of a frequently overlooked construct. *Academy of Management Review*. 5, 391–397 (1980). https://doi.org/10/c9zn4d.
39. Kerstholt, J.H.: The effect of time pressure on decision-making behaviour in a dynamic task environment. *Acta psychologica*. 86, 89–104 (1994). https://doi.org/10/fp2g5p.